\documentclass[11pt,twocolumn]{article}

\usepackage{times}

\usepackage[labelfont=bf]{caption}

\usepackage[top=0.5in,bottom=0.75in,left=0.5in,right=0.5in]{geometry}

\usepackage{url}

\usepackage{subfig}

\usepackage{graphicx}

\usepackage{amsmath}

\graphicspath{{figures/}{plots/}{.}}

\usepackage[numbers]{natbib}
\usepackage[bookmarksopen,bookmarksnumbered,colorlinks=true,allcolors=blue]{hyperref}

\usepackage{xcolor}

\usepackage[compact]{titlesec}

\usepackage{helvet}
\usepackage{sectsty}
\allsectionsfont{\sffamily}

\usepackage{mathptmx}   

\usepackage{flushend}

\usepackage{framed}

\usepackage[strict]{changepage}

\usepackage{subfig}

\newcommand{\sysname}{Textiverse}

\begin{document}


\title{\sffamily \sysname: A Scalable Visual Analytics System for Exploring Geotagged and Timestamped Text Corpora}

\author{Caroline Berger,$^{1}$ Hanjun Xian,$^{2}$ Krishna Madhavan,$^{3}$ and Niklas Elmqvist$^{1}$\\
\scriptsize $^1$Aarhus University, Aarhus, Denmark;
\scriptsize $^2$Facebook, Menlo Park, CA, USA;
\scriptsize $^3$Microsoft Corporation, Redmond, WA, USA}

\date{\sffamily October 2023}

\maketitle

\begin{abstract}
    We propose \sysname{}, a big data approach for mining geotagged timestamped textual data on a map, such as for Twitter feeds, crime reports, or restaurant reviews.
    We use a scalable data management pipeline that extracts keyphrases from online databases in parallel.
    We speed up this time-consuming step so that it outpaces the content creation rate of popular social media.
    The result is presented in a web-based interface that integrates with Google Maps to visualize textual content of massive scale.
    The visual design is based on aggregating spatial regions into discrete sites and rendering each such site as a circular tag cloud.
    To demonstrate the intended use of our technique, we first show how it can be used to characterize the U.S.\ National Science Foundation funding status based on all 489,151 awards.
    We then apply the same technique on visually representing a more spatially scattered and linguistically informal dataset: 1.2 million Twitter posts about the Android mobile operating system.
\end{abstract}


\begin{figure*}[tbh] 
    \centering
    \includegraphics[width=\linewidth]{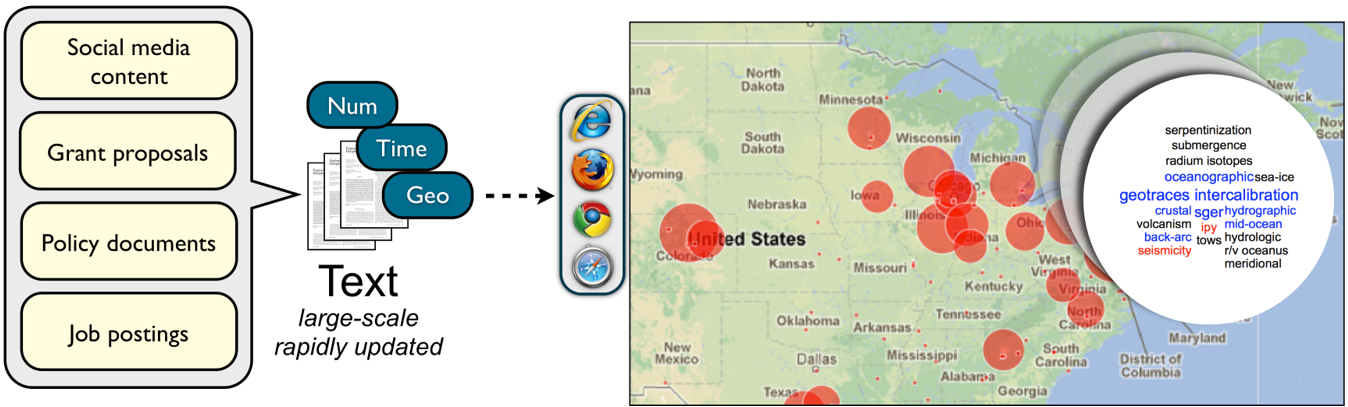}
    \caption{\textbf{\sysname{} for U.S. research awards.}
    Our \sysname{} visual analytics application being used to explore 489,151 U.S. National Science Foundation awards from 1976 to 2012.
    Highlighted is a snapshot of Woods Hole, MA in 2009.
    }
    \label{fig:teaser}
\end{figure*}
    

\section{Introduction}

Real-world datasets are often both complex and massive.
Representing such complex and large-scale data presents unique challenges in data mining and visual analytics. 
\textbf{Complexity} means that different data types each have their own ideal visual representations.~\cite{01-DBLP:journals/tog/Mackinlay86}
Techniques that are appropriate for visualizing quantitative data, textual, or relational data are often not applicable for the geospatial context. 
\textbf{Massive scale} gives rise to challenges such as data reduction, parallel computing, high-resolution displays, and user interfaces~\cite{02-DBLP:journals/cga/Ma01}.
Scientific visualization researchers have long treated scale as a core research problem, and have accordingly proposed a variety of architectures to enable high-speed data streaming,~\cite{03-DBLP:journals/cga/AhrensBMGLP01} adaptive compression and indexing,~\cite{04-Guoqing2011, 05-Shen2006} and parallel I/O.~\cite{06-DBLP:journals/cga/KendallHPLR11}
However, visual analytics and information visualization research has lagged behind, particularly in the context of increasingly popular web-based interfaces, where visually delivering a large volume of information is further limited by the browser capability and network bandwidth.
Also, a large dataset often implies frequent updates and rapid growth in data, which increases the difficulty and cost in preprocessing data.

In this paper, we propose a novel visual analytics system called \textsc{\sysname{}} for mining and visualizing large-scale, geotagged, and timestamped textual data.
Our implementation has a web-based visual interface and allows for exploring large multidimensional and multimodal datasets.
The intended use of \sysname{} is to explore enormous collections of text entries, each of which has a geospatial reference, a timestamp, and a numerical value.
To factor in these four attributes, we first plot markers of different sizes on a map to denote regional values.
Each marker can be turned into a circular dynamic tag cloud with the most significant tags in the center.
The temporal trends for geospatial sites are represented as animation of marker resizing.
Tag clouds can also be animated to show how some tags become more/less significant from one time period to another.
To view the temporal trends for a particular tag, we develop a similar design as SparkClouds~\cite{07-DBLP:journals/tvcg/LeeRKC10} to overlay a sparkline on each tag.

To address the scalability challenge for the web interface implementation, we extract keyphrases in parallel based on the full text and then aggregate them to reduce query execution time.
We use keyphrase extraction rather than traditional word-frequency solutions because phrases in general are more descriptive than single words.
However, keyphrasing algorithms usually consume significantly more computational resources than dealing with single words.
Also, we propose a refined \texttt{tf-idf} solution to tackle the frequently updating text.
Circular markers are loaded by default for browsing with a large number of nodes on the map at a time, whereas the interactive SVG versions provide interactive detail and are presented upon user request.
We describe in detail how we achieve this functionality given performance bottlenecks for processing large bodies of text. 

To demonstrate the typical use scenarios of our technique, we describe how it can be used to characterize the U.S.\ National Science Foundation (NSF) funding status based on all 489,151 awards over the years 1976 to 2012.
Each award has associated information such as the proposal summary, funding amount, awarded institutions, and awarded date, that correspond to the four dimensions in our design.
Our second application is used to visualize 1.3 million Twitter feeds about Android and iOS.
Finally, we discuss limitations of our technique and the potential of using it in other applications.


\section{Related Work}

Our work involves multidimensional, textual, and geospatial data, as well as large-scale data analytics.
Below we review relevant work in these fields.

\subsection{Multidimensional Visualization}

\textit{Multidimensional data} is often used interchangeably with \textit{multivariate} data in visual analytics research. 
However, when these two terms were first proposed, multidimensional data referred to data that had multiple independent parameters and their relationships, whereas multivariate focused on dependent variables~\cite{08-DBLP:conf/dbvis/BergeronCHKMTW93}. 
In Wong’s review~\cite{09-DBLP:conf/dagstuhl/WongB94}of 30 years’ multidimensional multivariate visualization (mdmv), he argued that such strict definitions had been discarded and both terms shifted towards a broader definition to study multiple variables regardless of their inter-dependency. 
The present study follows the modern and broad definition of these terms.

Among multidimensional visualizations, a line of research focuses on multiple variables but concerns only one or two data types.
Parallel coordinates~\cite{10-DBLP:conf/visualization/InselbergD90} and its extensions~\cite{11-DBLP:journals/cgf/ZhouYQCC08, 12-DBLP:journals/tvcg/Yuan2009, 13-DBLP:conf/visualization/FuaWR99} position variables as parallel axes and each data point is represented as a polyline that connects the corresponding points on each axis.
As a classic and effective multivariate visualization, parallel coordinate plots are widely used to visualize data sets that have multiple quantitative values.
Scatterplot matrices (SPLOMs)~\cite{DBLP:journals/tvcg/ElmqvistDF08} also aim to visualize multiple quantitative variables, but the approach is different.
The technique produces a collection of scatterplots between any two variables and organizes them in a matrix, where the diagram in a cell corresponds to the pairwise correlation between the row variable and the column variable.
Wong and Bergeron~\cite{15-DBLP:conf/visualization/WongB97} used the metric scaling technique to identify the inherent dissimilarities between quantitative attributes and accordingly reduced the data to low dimensionality.
Dust \& Magnet~\cite{16-DBLP:journals/ivs/YiMSJ05} uses a metaphor where each attribute is a magnet and each data point is a speck of iron dust that can be attracted or repelled by magnets.
All these efforts try to reduce multiple data dimensions to fit 2D displays.
However, they are all specifically designed for presenting quantitative variables and are therefore not capable of handling multiple data types.

Attempts have been made to visualize a data model that involves several disparate data types.
Chernoff~\cite{17-Chernoff1973} displayed cartoon-like faces on a map and used facial characteristics to represent a composite of nominal and quantitative values for demographics in a given geospatial regions.
Weber et al.~\cite{18-DBLP:conf/infovis/WeberAM01} visualized nominal and quantitative time-series data by developing a spiral visualization that plotted the time line as a spiral curve.
Different quantitative values, such as sunshine intensity, are mapped to the corresponding points on the spiral rendered in different colors, textures, line widths, or icons.
World Explorer~\cite{19-DBLP:conf/jcdl/AhernNNY07} helps users explore geo-referenced photos on Flickr with the map labelled with weighted tags associated with the photos.
GeoTime~\cite{20-DBLP:journals/ivs/KaplerW05} adds temporal information as Z index in a 3D environment to visually track events in an interactive view.
These studies integrate variables of different data types within a single view and generally offer more sophisticated interactions than single-type multidimensional visualizations.

When even more data types are needed in visualizations, existing approaches tend to rely on analysis and exploration to create multiple low-dimensional views that address different aspects of the data.
Polaris~\cite{21-DBLP:journals/tvcg/StolteTH02} (which later became the commercial tool Tableau) is a highly customizable visual interface for exploring large multidimensional data sets.
It offers table-based queries and a diversity of visualizations for identifying patterns and trends between two variables. 
Software libraries such as prefuse,~\cite{22-DBLP:conf/chi/HeerCL05} D3,~\cite{bostock2011d3} and the InfoVis Toolkit~\cite{24-DBLP:conf/infovis/Fekete04} require coding, but give users more freedom to explore multivariate data and provide many different kinds of visualizations to suit various data models.
However, the multiple views may cause loss of relationships between different contexts.
In this work, we aim to reconcile the need of multiple data types with the desire of maintaining contexts between views.


\subsection{Text Visualization}

Starting with the ubiquitous word clouds (or tag clouds) made popular by Flickr in 2001,~\cite{27-DBLP:journals/interactions/ViegasW08} text visualization is now widespread on the web.
The basic notion of text visualization is to summarize, highlight, and reduce potentially large bodies of text into compact visual representations, and has been a focus in information visualization research since its inception.~\cite{25-DBLP:conf/vl/Shneiderman96, 26-DBLP:conf/infovis/WiseTPLPSC95}
We identify three different types of text visualization: frequency-based, relational, and composite, and review them below.

\textit{Frequency-based techniques} focus on summarizing text primarily based on the frequency of words in a corpus.
Tag clouds~\cite{27-DBLP:journals/interactions/ViegasW08} is the canonical example, but several variations exist, such as Wordle,~\cite{28-DBLP:journals/tvcg/ViegasWF09} ManiWordle,~\cite{29-DBLP:journals/tvcg/KohLKS10} and EdWordle.~\cite{DBLP:journals/tvcg/WangCBZDCS18}

\textit{Relational text visualization} shows not just the content but also the context and relations of words and phrases.
Several examples exist:
WordTree~\cite{31-DBLP:journals/tvcg/WattenbergV08} is a visual and interactive concordance for phrases, DocuBurst~\cite{32-DBLP:journals/cgf/CollinsCP09} shows a document in the context of a word ontology, and Parallel Tag Clouds (PTCs)~\cite{33-DBLP:conf/ieeevast/CollinsVW09} visualize the relations between words and concepts in a document collection.

Finally, \textit{composite text visualization} often combines pure textual representations with other visualizations to convey additional data.
Examples include SparkClouds,~\cite{07-DBLP:journals/tvcg/LeeRKC10} which overlays a temporal trendline on the keywords in a tag cloud, TIARA,~\cite{34-DBLP:conf/kdd/WeiLSPZQSTZ10} ThemeRiver,~\cite{35-DBLP:journals/tvcg/HavreHWN02} TextFlow,~\cite{36-DBLP:journals/tvcg/CuiLTSSGQT11} context-preserving dynamic word clouds,~\cite{37-DBLP:journals/cga/CuiWLWZQ10} which integrate text with time-series charts, and WordBridge,~\cite{38-DBLP:conf/hicss/KimKEE11} which replaces the nodes and links in a graph visualization with node clouds and edge clouds that show the content of the relation.

The present study shares commonalities with composite text visualization, particularly with those combined with timely data. 
However, instead of focusing on generating topics and providing an additional graph, we present an animated tag cloud to represent change of tag significance over time.
Also, none of the above studies has attempted extremely large datasets, which as mentioned earlier require new solutions in data preprocessing and visual analytics.



\subsection{Thematic Map Visualization}

Thematic maps refer to a kind of maps that only use coastlines, boundaries, and places as points of reference for the phenomenon being mapped.~\cite{39-Thrower2007}
Visualization researchers have proposed a variety of techniques in representing a thematic map.
One solution is choropleth mapping,~\cite{41-DBLP:conf/iv/DangNS01} which aggregates data and renders regions based on pre-defined boundaries such as country and state.
Similar to choropleth map but without pre-defined boundaries, a dasymetric map~\cite{42-Eicher2001} determines the spatial divide based on the underlying data’s statistical distribution. 
A contour map~\cite{DBLP:conf/visualization/WijkT01, DBLP:conf/chi/AlperRH14} links data points of the same value with continuous smooth curves and is often used to describe 3D surfaces. 
These visualizations have been applied to demonstrate the geospatial distribution of numerical data, but have rarely been applied to multidimensional and textual data.

The proportional symbol technique has the potential to overlay more dimensions on a map. 
It communicates information associated with a location via the symbol content, shape, color, and size. 
Besides the capability of handling 1D numerical data~\cite{44-DBLP:journals/cartographica/JennyHH09} like the aforementioned techniques, it can represent composite datasets.~\cite{17-Chernoff1973, 45-Caquard2014} 
For example, GeoVISER supports exploration of thematic attributes and and geospatial data.~\cite{Yulong2022} 
The present study aims to propose a visual representation for multidimensional data and therefore, we choose the proportional symbol map as the geovisualization technique. 

\subsection{Large-Scale Visual Analytics}

Big data is the next frontier in computing.
Managing big data is one of the grand challenges of virtually any data-intensive computing discipline, visualization and visual analytics included.
Accordingly, much work is focused on this topic.

There are three major solutions to processing large-scale data.
The first approach is to reduce data volume without compromising data precision.
Pajarola~\cite{46-DBLP:conf/visualization/Pajarola98} and Shen~\cite{05-Shen2006} both used a hierarchical tree structure to encode and index scientific data at different granularity levels.
In information visualization, Guoqing et al.,~\cite{04-Guoqing2011} Wong et al.,~\cite{09-DBLP:conf/dagstuhl/WongB94} and Michaels et al.~\cite{47-Michaels1998} attempted to cluster variables according to their correlations and therefore reduced unnecessary attribute coupling.
In contrast to variable clustering, Fua et al.~\cite{48-DBLP:conf/visualization/FuaWR99} and Abello et al.~\cite{49-DBLP:journals/tvcg/AbelloHK06} focused on clustering data points that share similar characteristics among all the attributes.
Our study subsets a large dataset according to the zoom level and adaptively enables/disables interactions.
The second way is to build the visual analytics system using parallel computing techniques.
This technique is widely used in scientific visualization for rendering 3D objects,~\cite{50-DBLP:books/el/05/AhrensGL05, 51-DBLP:journals/pc/BeynonCCKSAFS02} but is less common in visual analytics.
In the present work, we attempt to deploy algorithms on a distributed computing environment for data transformation and visualization production.
The third and last solution is to cache and stream data on demand.
Ahrens et al.~\cite{03-DBLP:journals/cga/AhrensBMGLP01} discussed requirements for streaming scientific data: separable, mappable, and result invariant.
In the present work, we distribute individual visualization files to multiple servers.
These files contain all the information for a geographical region and are loaded only upon user request.

\subsection{Visual Analytics for Text}

Visual analytics systems provide a composite set of visual tools and weaves them together effectively to allow exploration of large-scale multivariate datasets.
Java et al.~\cite{52-DBLP:conf/kdd/JavaSFT07} and MacEachren et al.~\cite{53-DBLP:conf/ieeevast/MacEachrenJRPSMZB11} both propose temporal and geospatial visualizations based on Twitter data, but neither of them focuses on the textual content.
SentenTree visualizes frequent sequential text patterns and supports detail exploration of tweets through interaction~\cite{DBLP:journals/tvcg/HuWS17}.
Visual Backchannel~\cite{54-DBLP:journals/tvcg/DorkGWC10}, Spatiotemporal Social Media Analytics~\cite{55-DBLP:conf/ieeevast/ChaeTBJMEE12}, and I-SI~\cite{56-DBLP:journals/cgf/WangDMVCKR12} are visual analytic systems for analyzing social media data by topics, friendship, temporal trends, and geospatial distribution and can be used for event detection, topic exploration.
They incorporate extraction of topics, high-performance computing, and visual interface interactions into a system. 

The present study shares similar motivations of analyzing large-scale multidimensional data and adopts a similar approach of using geospatial visualization and text visualization techniques.
However, we propose a scalable solution for processing large-scale text corpora with a detailed performance evaluation for the preprocessing step.
Also, we present the evolution of topics using animated tag clouds. 

\begin{figure*}[tbh] 
    \centering
    \includegraphics[width=\linewidth]{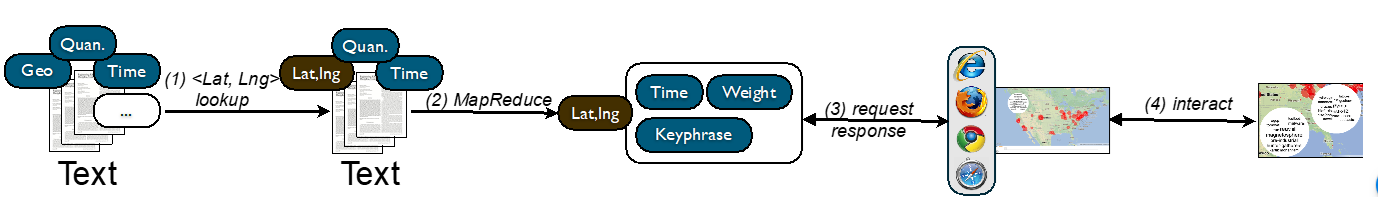}
    \caption{\textbf{\sysname{} system architecture.}
    An overview of the \sysname{} architecture.
    }
    \label{fig:system-architecture}
\end{figure*}

\section{\sysname: Large-Scale Spatial Text Analytics}

\textsc{\sysname} is a novel text visual analytics system to mine and visualize large-scale geotagged and timestamped textual data on a map.
The approach is based on incrementally extracting representative keywords and phrases from a large and dynamically updating dataset using parallel computation. 
The visual representation then aggregates spatial regions into discrete sites and renders each as a tag cloud. 
Tags are arranged in a circular layout with the most prominent text at center, and less significant ones towards the edge. 
From one time frame to another, the tag cloud is animated to reflect changes in text significance: increasingly significant tags are moved closer to the center, tags that experience a reduction in significance move to the edge, and new tags emerge with a ``phosphor'' effect.~\cite{DBLP:conf/uist/BaudischTCRHAZR06} 
To view the temporal trends of various tags, we overlay a sparkline on each tag. 

\subsection{Architecture}

\sysname{} takes any large-scale tabular-structure dataset such as CSV and SQL databases as input. 
As illustrated in Figure~\ref{fig:map-vis}, the dataset includes geospatial information, timestamp, and quantitative values for each textual unit such as a report, a Twitter feed, a job description, and a policy document. 
Other unused attributes are dropped and will not proceed to the next stage. 
Then we develop a mapper service to translate geospatial character strings into \texttt{$<$latitude, longitude$>$} values (step 1 in Figure~\ref{fig:system-architecture}). 
Next, we deploy and run a MapReduce job to aggregate input by latitude and longitude and extract keyphrases from each geospatial site (step 2). 
The resulted dataset is stored into a database, which can be queried by the web-based user interface (step 3). 
The web interface allows users to interact with the map and tag clouds by zooming, panning, and clicking so as to explore the timely trend, geographical distribution, and main topics in the dataset (step 4). 
The following sections present how the underlying infrastructure supports large-scale data processing and what the front end is capable of representing visually.

\begin{figure}[hbt] 
    \centering
    \includegraphics[width=\linewidth]{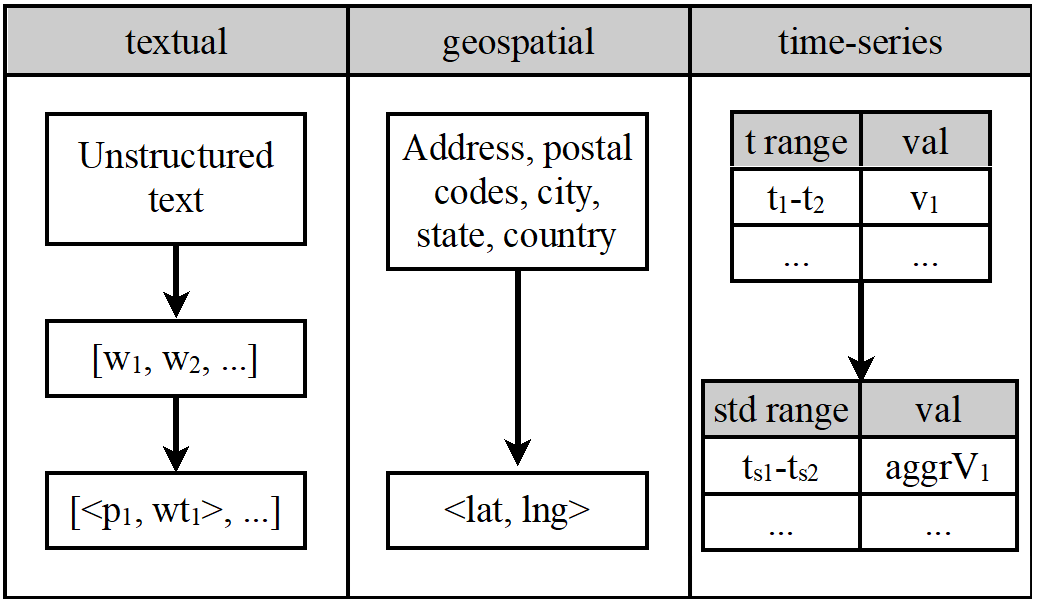}
    \caption{\textbf{\sysname{} data model.}
    Data model and preprocessing for \sysname{}.
    }
    \label{fig:data-model}
\end{figure}

\subsection{Data Management}

The input data for \sysname{} are text corpora, each of which is associated with a geospatial reference, a timestamp, and a quantitative value. 
Unstructured text is represented as a sequence of words with punctuations removed. 
There is no word limit or format requirement for a text corpus. 
All the word sequences are passed to our keyphrase extraction algorithm derived from GenEx.~\cite{57-DBLP:journals/corr/cs-LG-0212013} 
Unlike many tf-idf-based solutions that use all documents to be analyzed as the text corpus, we use a fixed third-party word-frequency database from American National Corpus (ANC). 
The ANC word-frequency statistics are based on a massive collection of generic written and spoken materials in American English. 
It contains 293,866 words and their frequencies with the top `the' having 1,081,168 occurrences and many uncommon words with only one occurrence. 
We adopt the ANC database to avoid reliance on target documents to be analyzed. 
As a result, newly updated documents will not be affected by the previous corpus and hence the order of processing documents does not have an effect on the final result. 
Also, newly updated documents do not influence earlier results and do not require a full rerun to obtain the most accurate result. 
This opens a new opportunity for parallelization and a significant reduction of cost for incremental changes. 
We discuss this further in the section on Scalability Considerations.

The basic idea of our algorithm is to (1) identify single keywords based on adapted \texttt{tf-idf} scores, (2) include words before or after those keywords to form phrases, (3) compute a phrase’s weight based on the sum of the \texttt{tf-idf} scores of all the words the phrase contains with an adjustment, (4) sort phrases by weight, and (5) select the top n phrases as keyphrases. 
Keyphrase extraction is selected against word-frequency because phrases in general are more descriptive and retain context of the original document better than single words. 
The resulted data set is an array of \texttt{$<$phrase, weight$>$} pairs. 
However, keyphrasing algorithms usually consume significantly more computational resources than those dealing with single words. 
This may become a performance bottleneck when processing a large corpus. 
We discuss this problem and other scalability issues below.

Geospatial data varies from address, postal code, to latitude-longitude pairs. 
We have implemented a preprocessing service to map geospatial data of various formats (such as ``Seattle, WA'', ``IN 47907'', ``Yellowstone National Park'') into latitude-longitude values.
This includes (1) mapping U.S.\ zip codes to latitude-longitude values based on the Zip Code Tabulation Areas (ZCTA) from the U.S.\ Census Bureau; (2) mapping international cities to latitude-longitude pairs using data from \url{http://GeoNames.org}; and (3) a general mapping of any address to coordinates using the Google Maps API.

Time-series data may be stored as a timestamp or a time range. 
To simplify, we model timestamps as a time range with the same start time and end time. 
A time range is divided into standardized units such as year, month, week, day, and hour. 
Quantitative values are distributed into these standardized units based on a function of time. Figure 3 illustrates the data management for the \sysname{} system.

After the preprocessing step above, input data are aggregated by geospatial location. 
As a result, for every spatial region, there is a list of \texttt{$<$std\_time\_range, phrase, weight$>$} triples. 
The weight here is adjusted based on the phrase weight and the quantitative value associated with a given time and a geospatial site. 

\begin{figure}[hbt] 
    \centering
    \includegraphics[width=\linewidth]{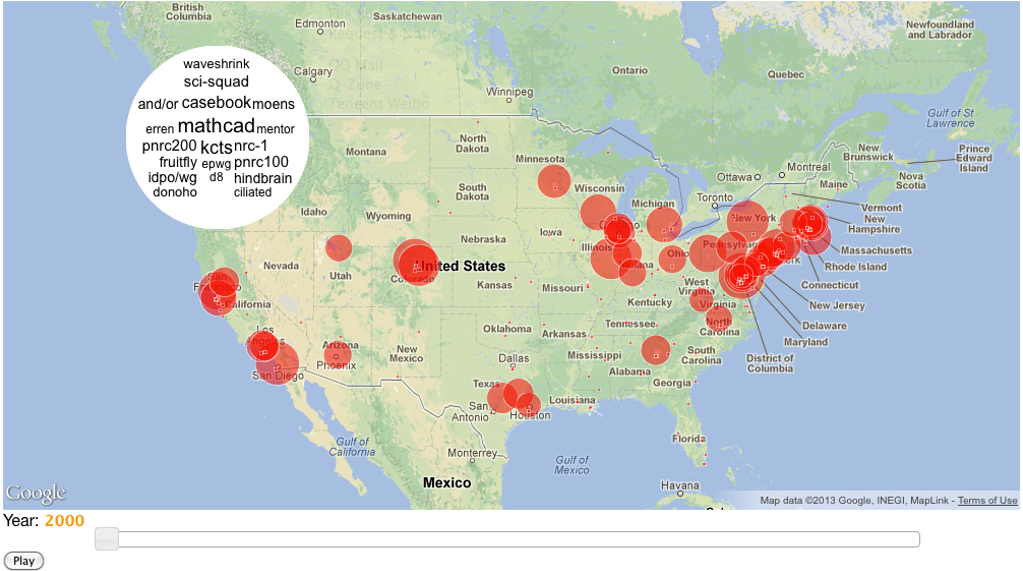}
    \caption{\textbf{\sysname{} map visualization.}
    Map view with less significant sites shown as small dots.
    }
    \label{fig:map-vis}
\end{figure}

\subsection{Animated Tag Clouds}

The tag cloud layout is derived from WordBridge,~\cite{38-DBLP:conf/hicss/KimKEE11} where the most significant tags are sized larger and placed in the center.
First, tags are sorted by weight and sized accordingly. 
Starting from the largest one placed in the center, it produces four available areas for the next tag. 
The next tag chooses the area closest to the center while large enough to fit the tag. 
The layout continues until all tags are displayed or no available areas are sufficiently large to hold tags.

Besides the above visual design, our layout algorithm also considers the potential movement of tags. 
The motivation is to make the layout stable as it is animated.  
This means that the tag cloud layout is also determined by its previous layout so that the tag movement is minimized when transitioning from one tag cloud to another. 
When several available boxes of free space can all fit a tag, the tag will be placed in the box that has the lowest sum of distance to the canvas center and distance from its prior position.

The time-series data associated with tags is rendered as a temporal trendline that overlays the text such as in SparkClouds.~\cite{07-DBLP:journals/tvcg/LeeRKC10} 
Similar to SparkClouds, the space limit makes it impossible to provide more details such as labels and legends on the chart. 
Therefore, the sparkline only gives a rough picture of the changes.

\subsection{Geospatial Layout}

Rendering tag clouds for all geospatial sites on a map is not only computationally infeasible~\cite{58-DBLP:conf/graphicsinterface/JohnsonJ08} but also produces overplotting (text overlap) that makes it difficult to interpret.
Geospatial visualizations often render nodes in different colors, shapes, and sizes to represent different data attributes. 
\sysname{} follows the same rule and thus must manage overplotting in dense regions just like any other geospatial visualization. 
One common solution is spatial clustering, which aggregates data within a certain geographical distance and represents them as one point. 
However, it has been reported~\cite{59-Slocum2022} that such data aggregation misleads users about the actual location that the data originate from. 
To avoid this, our approach plots nodes in their precise locations and reduces less significant points to small marks so as to alleviate the occlusion problem, as shown in Figure~\ref{fig:map-vis}. 
In the world view, we allow markers to occlude because markers at this zoom level convey the quantitative meaning more than delivering textual contexts. 
As users zoom in, markers are magnified at a lower rate than the map so that occluded ones start to separate.

A marker is expanded to show a tag cloud corresponding to the given time and location when a user clicks on the marker. 
Figure~\ref{fig:map-vis} demonstrates the tag cloud that overlays Seattle, WA in 2000. 
As a different time period is selected, all markers are animated to the new sizes and tag clouds are updated as discussed in the earlier section.

\subsection{Scalability Considerations}

Our goal with \sysname{} is to update the dataset for the web-based data explorer in a very short time when new input data are given. 
There are two major performance bottlenecks in this process: keyphrase extraction and network traffic between server and client.
As mentioned earlier, our keyphrase extraction algorithm is derived from GenEx~\cite{57-DBLP:journals/corr/cs-LG-0212013} but we do not use the analyzed documents as the text corpora for frequency calculation, but instead the ANC corpus. 
Therefore, there is no limit regarding the number of documents to be analyzed at a time or size of text assigned to the algorithm. 
There is also no difference in extraction results given different execution order. 
This brings a huge advantage in parallelizing keyphrase extraction because we can divide a job and distribute them to as many machines as possible without compromising the extraction quality. 
This is not like traditional \texttt{tf-idf} approaches, where the accuracy is improved as the number of documents increases. 
Also, big data often imply frequently updated records. 
When 100 new documents are to be added to the existing one billion documents, \texttt{tf-idf} solutions either extract tags from these new documents based on the existing large corpus or require a re-construction of statistics of all documents including the new ones. Such incremental updates again do not cause the same problem in our approach. 

Our solution adopts the MapReduce paradigm to parallelize keyphrase extraction in documents. 
First, \texttt{n} documents are distributed evenly to \texttt{k} machines. 
On each machine, our mapper takes the assigned documents and the ANC database as input. 
Each document is then divided into sentences and finally words. 
Second, each word’s weight is computed based on its local frequency and its rank in ANC. 
Similar to \texttt{tf-idf}, a high local frequency in a given document and a low rank in ANC indicate a high significance. 
Phrases that contain more words have a higher adjusted weight to offset the lower probability of occurrence. 
The mapper output is a key-value list where latitude-longitude values are keys and other fields (tag, weight, time) are values. 
Third, our reducer aggregates the mapper output to coordinates and stores the result in a database.

The second performance bottleneck is the network traffic required for transferring data to produce tag clouds and geographical information. 
We apply drill-down and caching techniques in presenting large-scale data. 
First, only the most significant sites are displayed and animated on the map while others are shrunk to a pixel. 
As users zoom in to a smaller region, one-pixel sites are invoked to regular sizes. Second, animated tag clouds are produced only upon user request. 
Third, latitude, longitude, and quantitative values are truncated to a less precise form when the zoom level is low. 
Last but not least, the animation duration can be used to preload data for the next possible action.

\section{Implementation}

The intended use of the \sysname{} visual analytics system is for very large datasets. 
Therefore, our implementation contains both server-side and client-side components. 
The server-side component uses a scalable data management pipeline that extracts keyphrases from online databases in a Hadoop cluster. 
We deploy our keyphrase extraction program in a Hadoop cluster that has about 800 nodes, each with two 2.33 GHz Quad-core Intel E5410 CPUs and 16GB memory. 
The keyphrase extraction algorithm is implemented in Python and it reads input data from a MySQL database. 
Each node processes an even number of documents and stores the extracted keyphrases and weights in a CSV file and finally into a database. 

The web-based implementation is built in JavaScript and uses the Google Maps API to visualize textual content of massive scale. 
The client requests data from a JSON-RPC server implemented in PHP and renders the map and markers using Google Maps. 
The tag cloud is produced as an SVG object using RaphaelJS and is overlaid on top of the map. 
The SVG format is selected for its increasing popularity in web-native information visualization [58]. 
On average, each map update with top 200 sites needs to acquire 2.4KB data from the server and each tag cloud update needs about 850 Bytes.

\section{Example 1: NSF Grant Data}

To demonstrate the use of \sysname{}, we here give an example based on 489,151 U.S.\ National Science Foundation awards over the period 1976 to 2012. 
This data set is publicly available on \url{http://nsf.gov/} and it fits the desired data model for \sysname{}: textual data are proposal abstracts with on average 1,829 characters per abstract; each award is associated with an institution or a company, mostly with a detailed address and a zip code; and any grant has an active period.

\begin{figure}[hbt] 
    \centering
    \includegraphics[width=\linewidth]{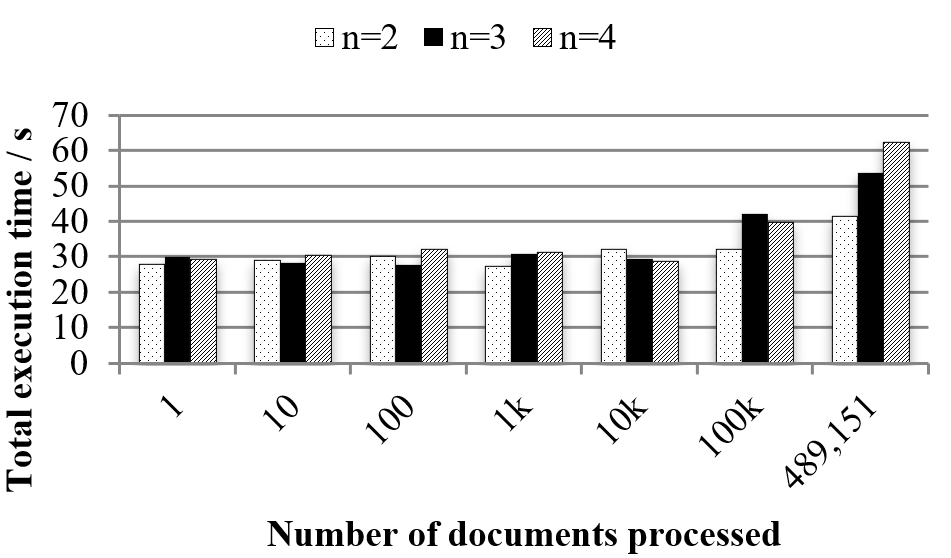}
    \caption{\textbf{Example 1 performance.}
    Performance of keyphrase extraction algorithm (Hadoop).
    }
    \label{fig:performance1}
\end{figure}

We run our keyphrase extraction algorithm on proposal abstracts to compute up to 4 keyphrases (with a maximum of four words in a phrase) per abstract.
The weight of a keyphrase is determined by the adapted tf-idf score, the adjustment based on the number of words in a phrase, and the grant’s award amount are aggregated by institution (we assume that each institution has a unique zip code).

\begin{figure}[hbt] 
    \centering
    \includegraphics[width=\linewidth]{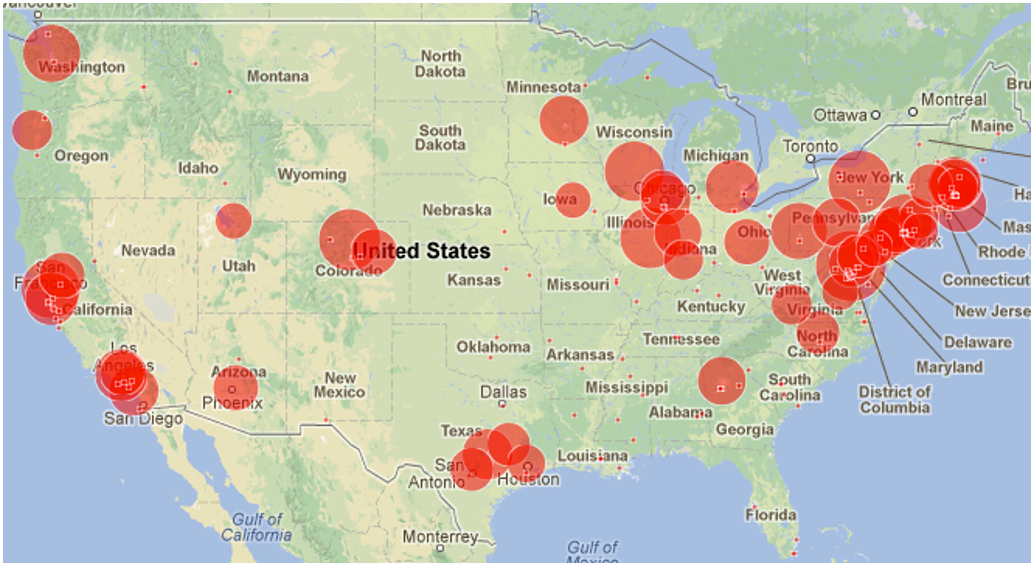}
    \caption{\textbf{NSF investments.}
    An overview of NSF investments in the U.S. in 2008.
    }
    \label{fig:nsf-vis1}
\end{figure}

We measure the execution time of our keyphrase extraction algorithm using 179 nodes in Hadoop to process 1, 10, 100, to 100k, and finally all 489,151 NSF award abstracts. 
The documents are randomly selected from the NSF award data. 
We further test the difference between 2-gram, 3-gram, and 4-gram extraction. 
An n-gram keyphrase extraction refers to the maximum number of words allowed to form a phrase. 
For instance, a 3-gram algorithm may recognize ``information visualization technique'' as a keyphrase, while a 2-gram algorithm cannot. 
A higher n consumes more computational resource but is able to identify longer phrases. 
As shown in Fig.~\ref{fig:nsf-vis1}, the execution time remains almost the same for processing 10,000 or fewer documents. 
For 100k and above, the execution time increases slowly, with the longest processing time of 62.4 seconds for 4-gram extraction of all 489,151 awards. 
Also in Fig.~\ref{fig:nsf-vis1}, the difference in n has a very limited effect on execution time when the number of documents is 100k or fewer. 
The time cost of processing only one document (about 26 seconds) is a clear indicator of the total overhead of starting/quitting the program, Hadoop scheduling, database I/O, file I/O, and building hashmaps for the ANC database. 
These efforts are mandated regardless of how many documents are processed. 
We also implement a sequential program following the same algorithm and the execution time for processing all 489,151 documents is 6,172 seconds (almost 100 times longer). 
The sequential approach has the time advantage for processing less than 1,000 documents because it does not need to pay the overhead of the map-reduce procedure. 
However, as more documents are added, the execution time of the sequential solution increases almost linearly with the number of documents.

\begin{figure}[hbt] 
    \centering
    \includegraphics[width=\linewidth]{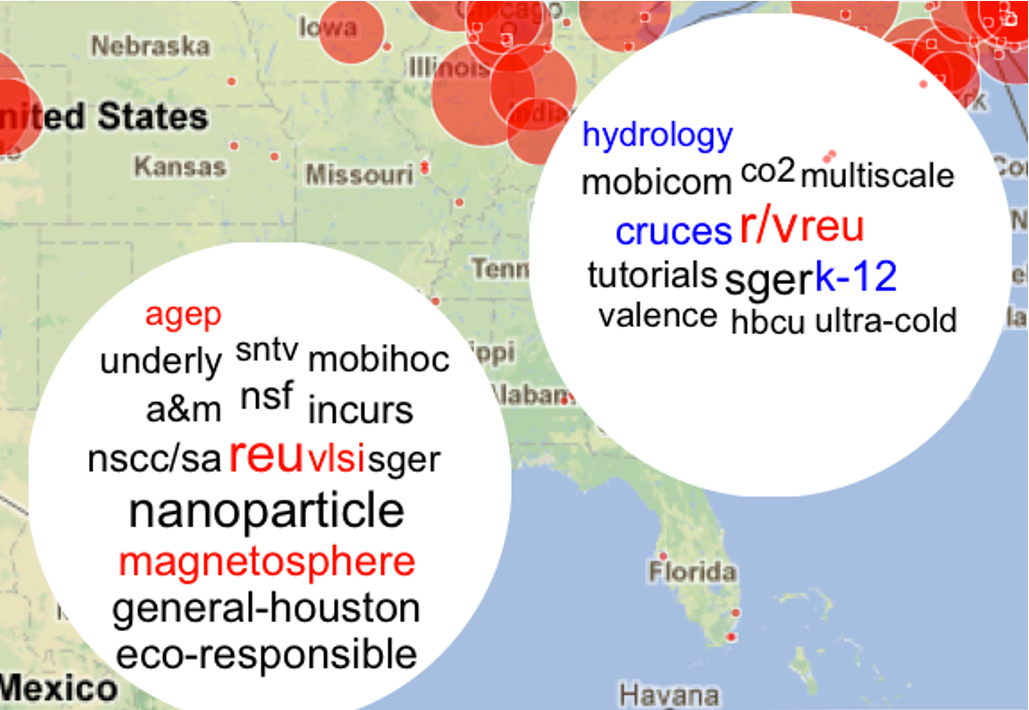}
    \caption{\textbf{NSF investment changes.}
    Changes of funded research topics in 2009.
    }
    \label{fig:nsf-vis2}
\end{figure}

Regarding the user interface, this example uses circle size to denote the numerical value associated with each location. 
The total award amount received by an institution is presented on a log scale. 
Top 200 sites are returned for the user’s current view port.
Fig. 6 shows an overview of the funding distribution across institutions in the U.S.\ in 2008. 
Based on the location of prominent nodes, regions near New York City, Washington D.C., Chicago, San Francisco, San Diego, Houston/San Antonio, Seattle, and Denver all seem to have institutions that receive significant amounts of NSF funding. 
Suppose a user is interested in what NSF has been investing in Texas and North Carolina; to investigate further, they click the two markers and open the tag clouds, as shown in Fig.~\ref{fig:nsf-vis2}. 
It becomes obvious that institutions in Houston have magnetosphere and REU as their primary research foci, whereas Duke University played an important role in polar research (R/V). 
From 2008 to 2009 in Fig.~\ref{fig:nsf-vis2}, a new `nanoparticle' suddenly becomes the most prominent topics in Houston and magnetosphere and REU remains important but not as much as in 2008 (indicated by red).
On the other hand, K-12 research becomes more significant in Duke University (shown in blue).
\begin{figure}[hbt] 
    \centering
    \includegraphics[width=\linewidth]{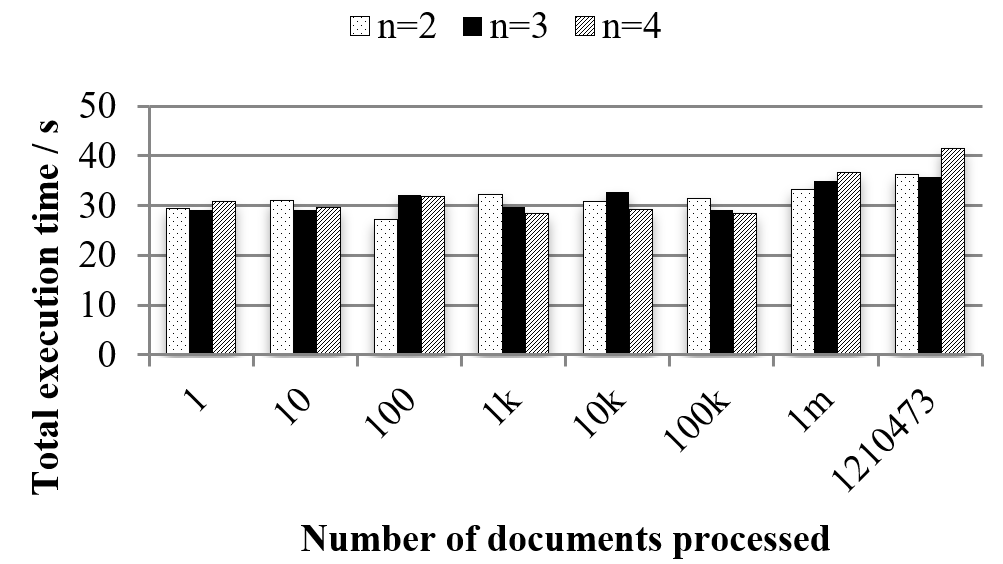}
    \caption{\textbf{Example 2 performance.}
    Performance of keyphrase extraction algorithm (Hadoop). 
    }
    \label{fig:performance2}
\end{figure}

\section{Example 2: Twitter Feeds}

\begin{figure}[hbt] 
    \centering
    \includegraphics[width=\linewidth]{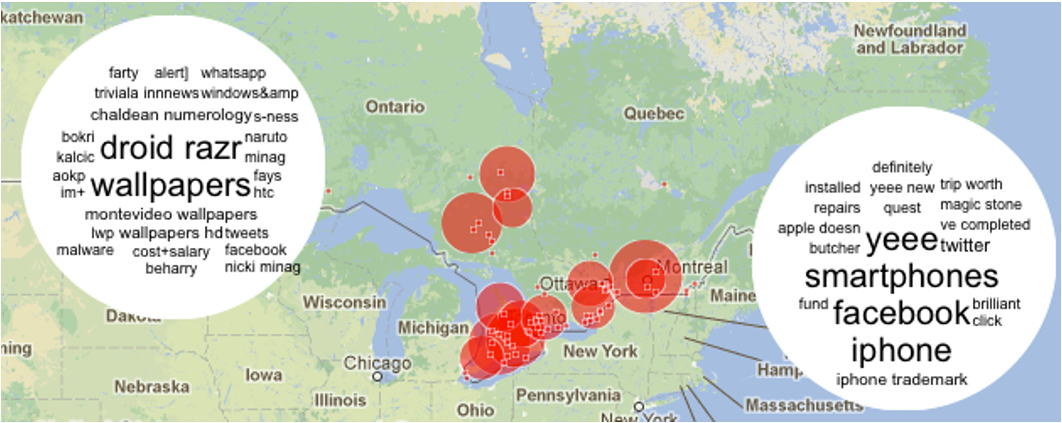}
    \caption{\textbf{Android reviews in Canada.}
    Comparison of different topics related to Android in two cities in Canada during March 2013.
    }
    \label{fig:android-vis}
\end{figure}

Our second experiment applies \sysname{} to show 1.2 million Twitter feeds about Android published during March 2013. 
Compared to NSF award data, Twitter feeds shares the same multidimensional and large-scale nature. 
Tweet content, timestamp, and geotags are also available and therefore fit the data model of \sysname{}. 
However, tweets are often shorter in length, come from a wider range of geospatial locations, and include a large number of informal language and external links. 
Therefore, we adjust a few parameters for this scenario such as reducing the number of keyphrases extracted to no more than three, introducing the spoken American English database from ANC, and removing URLs from the tweets. 
Among all the tweets, only English-written ones are included in this example. 
If a tweet contains non-ASCII characters or has more than half of its words that cannot be matched to any word in ANC, it is considered as non-English. 
Twitter feeds are recognized as related to Android if `Android' appears at least once in the tweet content. 
Since the geotags included in tweets are formatted as city, state, and country, we use a different address to latitude-longitude mapper to translate it into coordinates. 
Tweets without geotags or timestamps are excluded in this example. 
Because the tweet content and sentences are much shorter than NSF awards, the total execution time for extracting keyphrases from 1,210,473 tweets is only 1,219 seconds for the sequential algorithm and 41.6 seconds for the parallel one when deployed in the same Hadoop cluster with the same configuration (n=4). 
Again, we subset tweets to different numbers (such as 1, 10, 100, ..., 1 million, and all 1,210,473) and similar results as the NSF case have been found, as shown in Fig.~\ref{fig:android-vis}. 
The execution times for processing 1 million or fewer tweets are largely influenced by the minor variations in overhead. 
A slight increase in time is observed for analyzing more Twitter feeds. 
Also, the difference in how many words per phrase (n-gram) contributes insignificantly to the total execution time. 
This means, the cluster and technique used in this paper are able to outpace the new tweet updates.

In this example, the mark size indicates the number of tweets and each tweet is weighted equally. 
Fig.~\ref{fig:android-vis} shows the main topics discussed related to Android in March 2013 in two cities in Canada. 
Although Twitter is a social networking site without any geographical barrier in communication, the two cities differ significantly in their major topics. 
The left one has more conversations on a particular cell phone model and wallpapers, whereas Twitter users in the right site are more interested in discussing smartphones in general, comparing Android devices with iPhone, and sharing Facebook news/links.

\section{Conclusion and Future Work}

In this paper, we propose a novel visual analytics application called \sysname{} for mining and representing geotagged timestamped textual data on a map. 
Spatial regions are aggregated into discrete sites and rendered as animated tag clouds on a map. 
Because our intended use of the \sysname{} technique is for very large datasets, we present a practical web-based implementation of the tool that integrates with Google Maps to visualize textual content of massive scale. 
Also, we discuss the advantage of an adapted keyphrase extraction algorithm to increase parallelization. 
Our solution of using Hadoop in extracting keyphrases from a large number of documents obtains a huge performance gain when processing a large number of texts. 

To demonstrate the intended use of our technique, we show how it can be used to characterize NSF funding status based on all 489,151 awards over the year 1976 to 2012, as well as popular mobile topics worldwide based on 1.2 million Twitter feeds in March 2013. 
There are many other potential usages for \sysname{}. 
For example, \sysname{} can be used to demonstrate the geographical distribution and timely trend of the U.S.\ job market based on positions posted on LinkedIn, Monster, and Indeed. 
Another possible application of \sysname{} is to compare and review policies based on policy documents issued by different countries, states, and cities. 
Also, it can be used to study how the same product is adopted and regarded differently around the world by analyzing customer reviews.

\section*{Acknowledgments}

This work was partially supported by the U.S.\ National Science Foundation grant TUES-1123108.
Any opinions, findings, and conclusions, or recommendations expressed here are those of the authors and do not necessarily reflect the views of the funding agencies.

\bibliographystyle{plainnat}
\bibliography{textiverse}

\end{document}